\documentclass[fleqn,usenatbib]{mnras}

\usepackage{newtxtext,newtxmath}

\usepackage{multirow}
\usepackage{xspace}
\usepackage{graphicx}	
\usepackage{amsmath}	
\usepackage{amssymb}	
\usepackage{hyperref}
\usepackage[capitalise]{cleveref}
\usepackage[normalem]{ulem}

\newcommand{\degsq}{\, \mathrm{deg}^2}
\newcommand{\E}{\times10}
\renewcommand{\d}{{\rm d}}
\newcommand{\thetaline}{\boldsymbol{\theta}_{\rm line}}
\newcommand{\thetashape}{\boldsymbol{\theta}_{\rm shape}}
\newcommand{\model}{\mathcal{M}}
\newcommand{\noise}{\mathcal{N}}
\newcommand{\data}{\mathcal{D}}
\newcommand{\params}{\boldsymbol{\theta}}
\newcommand{\skadssax}{S3-HI\xspace}
\newcommand{\skadssex}{S3-Cont\xspace}
\newcommand{\lnb}{\ln(B) > 0}
\newcommand{\snrvel}{\mathrm{SNR_{vel}} > 5}
\newcommand{\multinest}{\textsc{MultiNest}\xspace}
\newcommand{\hi}{H\,\textsc{i}\xspace}

\newcommand{\ml}[1]{{#1}}

\newcommand{\jbca}{{Jodrell Bank Centre for Astrophysics, School of Physics \& Astronomy, The University of Manchester, Manchester M13 9PL, UK}}
\newcommand{\skasa}{{SKA SA, The Park, Park Road, Cape Town 7405, South Africa}}
\newcommand{\aims}{{African Institute for Mathematical Sciences, 6 Melrose Road, Muizenberg, 7945, South Africa}}
\newcommand{\ucl}{{Department of Physics and Astronomy, University College London, Gower Street, London WC1E 6BT, UK}}

\newcommand{\myemail}{ian.harrison-2@manchester.ac.uk}
\newcommand{\michelleemail}{michelle@ska.ac.za}

\title[Bayesian H\,{\normalsize \it I} line redshifts]{Redshifts for galaxies in radio continuum surveys\\from Bayesian model fitting of H\,{\Large \bf I} 21-cm lines.}
\author[Harrison, Lochner \& Brown]{
Ian Harrison,$^{1}$\thanks{\myemail}
Michelle Lochner,$^{2,3,4}$\thanks{\michelleemail}
Michael L. Brown$^{1}$
\\
$^{1}$\jbca\\
$^{2}$\aims\\
$^{3}$\skasa\\
$^{4}$\ucl
}

\date{Accepted XXX. Received YYY; in original form ZZZ}

\pubyear{2017}

\begin{document}
\label{firstpage}
\pagerange{\pageref{firstpage}--\pageref{lastpage}}
\maketitle

\begin{abstract}
We introduce a new Bayesian \hi spectral line fitting technique capable of obtaining spectroscopic redshifts for millions of galaxies in radio surveys with the Square Kilometere Array (SKA). This technique is especially well-suited to the low signal-to-noise regime that the redshifted 21-cm \hi emission line is expected to be observed in, especially with SKA Phase 1, allowing for robust source detection.  After selecting a set of continuum objects relevant to large, cosmological-scale surveys with the first phase of the SKA dish array (SKA1-MID), we simulate data corresponding to their \hi line emission as observed by the same telescope. We then use the \multinest nested sampling code to find the best-fitting parametrised line profile, providing us with a full joint posterior probability distribution for the galaxy properties, including redshift. This provides high quality redshifts, with redshift errors $\Delta z / z <10^{-5}$, from radio data alone for some $1.8\E^6$ galaxies in a representative $5000 \, \deg^2$ survey with the SKA1-MID instrument with up-to-date sensitivity profiles. Interestingly, we find that the SNR definition commonly used in forecast papers does not correlate well with the actual detectability of an \hi line using our method. We further detail how our method could be improved with per-object priors and how it may be also used to give robust constraints on other observables such as the \hi mass function. We also make our line fitting code publicly available for application to other data sets.
\end{abstract}

\begin{keywords}
large-scale structure of Universe --- radio continuum: galaxies  --- techniques: spectroscopic
\end{keywords}



\section{Introduction}
\label{sec:introduction}
The Square Kilometre Array (SKA)\footnote{\url{http://www.skatelescope.org}} will perform the kind of deep, wide surveys which are capable of delivering world-leading cosmological constraints from radio wavelengths using probes including galaxy clustering \citep{2015MNRAS.450.2251Y}, weak gravitational lensing \citep{2016arXiv160103947H}, \hi intensity mapping \citep{2015ApJ...803...21B} and ultra-large scale tests of general relativity and Gaussianity \citep[e.g.][]{2015MNRAS.448.1035C}. However, for those probes which require redshifts for individual sources, good redshift estimates may be difficult to obtain. The emission mechanism for the ordinary star-forming galaxies expected to form the bulk of sources in SKA cosmology catalogues is from synchrotron which has a uniform spectral slope of $-0.7$ across a large frequency range. This means the galaxies' spectra as measured by SKA will be almost completely featureless, with redshift and flux entirely degenerate over the relevant frequency range. The expectation in previous analyses has been that redshifts could be obtained in two ways: spectroscopic redshifts from high-significance detections of the \hi 21-cm line emission from the sources, or photometric redshifts from cross-matching the radio continuum sources with overlapping surveys at optical and near-infrared (nIR) wavelengths. However, the number of catalogue matches may be small (as found in \citealt{2010MNRAS.401.2572P, 2016MNRAS.456.3100D, 2016MNRAS.463.3339T}, though see also the larger matching fractions found in \citealt{2017arXiv170309719S}), and photometric redshifts may be subject to significant uncertainties, biases and catastrophic outliers, so it will be important to extract as much information as possible from the \hi 21-cm line. The lack of redshift information for e.g. tomographic binning of sources in weak lensing is a potentially limiting factor on the cosmological power of SKA surveys.

Here we investigate the ability of a Bayesian model-fitting approach to estimate the redshift of radio continuum sources using the \hi 21-cm emission line and apply the technique to simulated data from the first phase of the SKA mid-frequency dish array (SKA1-MID). We use the continuum detection of a galaxy as prior information, reducing the redshift estimation problem to fitting a six parameter model to a one dimensional data set. A similar approach has been implemented by \cite{2012MNRAS.423.2601A, 2012PASA...29..221A}, who fit absorption line profiles using Gaussian mixture models. For emission line fitting, the \texttt{SOFIA} package \citep{2015MNRAS.448.1922S} performs source finding using threshold methods on full three dimensional data cubes.

Using the technique described here, we find high quality \hi emission line redshifts (with high spectroscopic precisions and with outlier fractions at least as good as typical photometric redshift methods) may be obtained for around $10$ per cent of the star-forming galaxies in a fiducial SKA1 continuum cosmology survey. We compare the catalogue resulting from this detection algorithm to that resulting from the Signal-to-Noise Ratio (SNR) definition and cut from  \cite{2015MNRAS.450.2251Y,2015aska.confE..21S}. We find a comparable number of sources, with a comparable redshift distribution, remain when the full detection algorithm is simulated, although note that the exact set of sources differs \ml{significantly} between the two catalogues.

This paper is organised as follows: in \cref{sec:method} we introduce the \ml{Bayesian} method used to fit a six parameter model to the \hi line data; then in \cref{sec:data} we describe the creation of the simulated data catalogue from the \cite{2009ApJ...703.1890O} \skadssax simulation and the relevant observation parameters for SKA1-MID surveys in Band 1 and Band 2. \cref{sec:results} then describes the results of this procedure, with \cref{sec:extensions} detailing potential improvements and extensions and \cref{sec:conclusions} describing our conclusions.

We also provide our code for simulation of SKA1-MID \hi line catalogues, and our analysis code at \url{http://github.com/MichelleLochner/radio-z} for application of our method to other simulations and data and to enable comparisons of the performance other methods on our simulated data set.

\section{A brief introduction to Bayesian statistics}
\label{sec:bayes}
The problem of spectral line fitting has two components: one is to robustly determine whether or not a spectral line is present, the other is to then find the best fitting parameters of the line model and their uncertainties. Both of these can be done elegantly within a Bayesian statistics framework. We therefore give here a very brief introduction to Bayesian statistics, referring the reader to (for example) \cite{trotta} for a more in-depth review.

Bayes' theorem is given by:
\begin{equation}
P(\params|\data) = \frac{P(\params) P(\data |\params)}{P(\data)},
\end{equation}
where $\data$ represents the data and $\params$ represents the vector of parameters for the chosen model. $P(\params)$ is called the \emph{prior} and is the probability distribution of the parameters, before any data is taken. This is usually derived from physical constraints for the problem at hand, for example ensuring a density parameter remains positive, but can also include constraints from previous experiments. $P(\data|\params)$ is the \emph{likelihood}, the probability of the data, given a set of values for $\params$. This informs how likely a given set of parameters is, in light of the data at hand. $P(\params|\data)$ is the \emph{posterior} and is generally the quantity scientists are interested in: what is my degree of belief in a chosen theory (with given parameters), now that I have taken some new data? Lastly, $P(\data)$ is called the \emph{evidence} or marginal likelihood and is a normalisation constant that is crucial to model selection (see below), but unimportant for parameter inference. It can thus be seen that Bayes theorem is prescription of how to update one's degree of belief in a particular theory, using a new set of data.

Bayesian inference then proceeds to determine the full posterior over all parameters in the model. The best fitting values for the parameters are those that maximise the posterior. However to determine the uncertainty on each parameter, all other parameters must be \emph{marginalised} over to produce a one-dimensional posterior. This marginalisation results in an integral over parameter space:
\begin{equation}
P(\phi|\data) = \int P(\phi, \boldsymbol{\psi}|\data) d\boldsymbol{\psi},
\end{equation}
where $\phi$ is the parameter of interest and $\boldsymbol{\psi}$ is the vector of remaining parameters to be marginalised over. In the vast majority of problems, this integral cannot be solved analytically. Fortunately, several numerical techniques exist, including Markov Chain Monte Carlo \citep{Metropolis1953, Hastings1970} and Nested Sampling \citep{skilling2006}, that make Bayesian parameter inference tractable.

Bayesian statistics also provides a framework in which to perform robust model selection. The important quantity for this is the evidence which is computed, for a given model $\model$, by integrating over the entire parameter space:
\begin{equation}
P(\data|\model) = \int P(\data | \params, \model) P(\params|\model) d\params.
\end{equation}
The evidence naturally incorporates an Occam's razor effect, penalising models with large prior volumes (for example with many parameters) unless they provide a significantly improved fit to the data. 

The evidence is used in model selection when comparing two models, $\model_1$ and $\model_2$, by computing the ratio of posterior odds (which is simply further application of Bayes' theorem):
\begin{equation}
\frac{P(\model_1|\data)}{P(\model_2|\data)} = \frac{P(\data|\model_1)}{P(\data|\model_2)} \frac{P(\model_1)}{P(\model_2)}.
\end{equation}
Given that there is usually no strong reason to a priori prefer one model over another, the model priors $P(\model_1)$ and $P(\model_2)$ are often set to be equal, making the most important quantity the ratio of evidences for each model. This is known as the Bayes factor:
\begin{equation}
\label{eqn:bayesfactor}
B = \frac{P(\data|\model_1)}{P(\data|\model_2)}.
\end{equation}
The Bayes factor can be directly used to select one model over another by simply comparing if the evidence is greater for one over the other. The Jeffrey's scale \citep{jeffreys} can be used to decide how strong the evidence is for one model over another, where $\rm{ln}(B)>5$ constitutes strong preference for $\model_1$ and $\rm{ln}(B)>0$ weak.

\section[]{Bayesian model fitting for H\,{\sevensize\bf I} line profiles}
\label{sec:method}
We now illustrate how Bayesian statistics can be used to solve the \hi line detection and characterisation problem. We consider the case in which a galaxy has been detected in a radio continuum image, providing information on its sky location. This provides us with prior information on spectral line data: rather than searching a full image cube for the emission line from this galaxy, we may search a one dimensional data vector corresponding to this sky location (with filtering applied to leave the source spatially unresolved) and consisting of a series of flux measurements as a function of frequency $\data(\nu_{i})$. We further assume that the \hi 21-cm emission from the galaxy can be well modelled by the six parameter double horn profile $\Psi(\nu | z, \Psi^{\rm obs}_{\rm max}, \Psi^{\rm obs}_{0}, w^{\rm obs}_{\rm peak}, w^{\rm obs}_{50}, w^{\rm obs}_{20})$ described in \cite{2009ApJ...703.1890O}, an example of which is shown in \cref{fig:example_line}. The two $\Psi^{\rm obs}$ parameters give the line heights at the maximum (i.e. the peak of the horns) and the centre of the line (at the bottom of the dip between the horns) respectively, and the three $w^{\rm obs}$ parameters give the width of the line (which is a function of the galaxy's inclination angle and rotational velocity) at 100, 50 and 20 per cent of the peak height. We also assume here that the continuum component of the emission has been successfully fully removed, meaning we are only fitting the \hi line emission. We do however note that our method could be readily extended to simultaneously fit a more complicated model including continuum or more complicated line profiles such as the `Busy function' described in \cite{2014MNRAS.438.1176W}.

With the full six parameters defined as $\thetaline$ and the latter five (i.e. excluding $z$) as $\thetashape$, we then use the nested sampling code \multinest \citep{2009MNRAS.398.1601F, 2013arXiv1306.2144F} to map the full joint posterior distribution $P(\thetaline | \data, \model)$ given the data vector and model hypothesis $\model$ that there is an emission line present in the data. The redshift probability distribution for each source is then given by the marginalisation of this posterior distribution over the five shape parameters:
\begin{equation}
P(z | \data, \model) = \int \d \thetashape \, P(\thetaline | \data, \model).
\end{equation}
Here we are not merely interested in the case where the \hi emission line is significantly detected above some threshold, providing a `spectroscopic' redshift with extremely narrow $P(z)$, but in the properties of the full set of $P(z)$ for all continuum detected sources. Even relatively broad $P(z)$ (such as those often provided by photometric methods at optical and infrared wavelenghts) can still provide extremely useful information in terms of redshift binning for cosmological surveys, being summed to provide an estimate of the redshift number distribution of sources $n(z)$. For the six fitted parameters we adopt broad, uninformative priors on the set of parameters we fit, detailed in \cref{tab:priors} and set by the range depicted in the full simulated input catalogue from \cite{2009ApJ...703.1890O}.
\ml{The python code takes around 10 minutes to run on a normal laptop for an average galaxy and is trivially parallelisable at the catalogue level.}

\subsection{Identifying false detections with the evidence}
Observing with the SKA (and indeed all radio telescopes) takes place within frequency bands of finite width. For sources which are detected in continuum but whose \hi 21-cm lines are redshifted to outside of the observing band the data vector $\data(\nu_{i})$ will contain only noise, and other faint sources will be buried deep within the noise. In order to attempt to remove spurious detections of emission lines in noise-only data, we make use of Bayesian model selection, as outlined in \cref{sec:bayes}. The two models we compare are the six-parameter \hi profile outlined above, $\model$, and a pure noise model, $\noise$, where the signal is consistent with zero. We compute the Bayes factor, $B$, using \cref{eqn:bayesfactor}, to eliminate spurious line detections and have control over the purity of the sample we produce.

In \cref{sec:results} we present results with a variety of cuts on $B$, in order to exclude spurious noise detections, but we emphasise such cuts are not strictly necessary, and that $B$ could be used as a weight factor in derived analyses which make use of the $P(z)$ obtained for all continuum sources, such as in estimating number density distributions $n(z)$.
\begin{table}
\caption{Prior shapes and ranges on fitted parameters used. $v_0$ priors for each band correspond to the edges of each observing band, apart from the upper limit in Band 2 which corresponds to the constraint that redshift must be positive.}
\label{tab:priors}
\begin{center}
\begin{tabular}{@{}clc}
\hline
Parameter & Prior Shape & Prior Range \\
\hline
$v_{0}$ (Band 1) & Uniform & $-226\E^{3}$ - $-78.2\E^{3}$ \\
$v_{0}$ (Band 2) & Uniform & $-99.3\E^{3}$ - $0$ \\
$w^{\rm obs}_{20}$ & Uniform logarithmic & $10^{-1}$ - $10^{7.5}$ \\
$w^{\rm obs}_{50}$ & Uniform logarithmic & $10^{-1}$ - $10^{7.5}$ \\
$w^{\rm obs}_{\rm peak}$ & Uniform logarithmic & $10^{-1}$ - $10^{7.5}$ \\
$\Psi^{\rm obs}_{\rm max}$ & Uniform logarithmic & $10^{-11}$ - $10^{-2}$ \\
$\Psi^{\rm obs}_{0}$ & Uniform logarithmic & $10^{-11}$ - $10^{-2}$ \\
\hline
\end{tabular}
\end{center}
\end{table}

\section{Simulated observational data}
\label{sec:data}
Before the advent of the SKA, a number of precursor and pathfinder telescopes will operate, with some performing \hi emission line surveys, such as LADUMA on MeerKAT \citep{2012IAUS..284..496H} and WALLABY on ASKAP \citep{2012MNRAS.426.3385D}. Here we consider the performance of SKA1 surveys, in the understanding that the simulated data and populations will be similar for precursor surveys (MeerKAT will be integrated into SKA-MID and shares similar noise performance on each individual antenna).
\begin{figure*}
\includegraphics[width=0.475\textwidth]{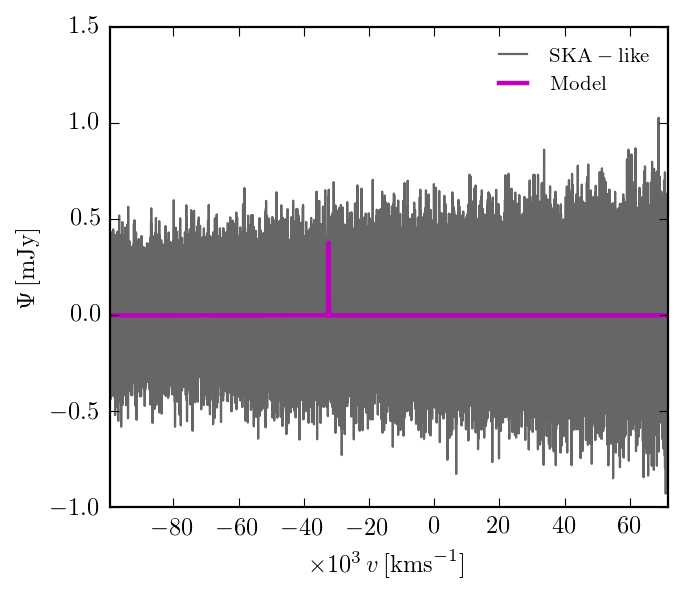}
\includegraphics[width=0.475\textwidth]{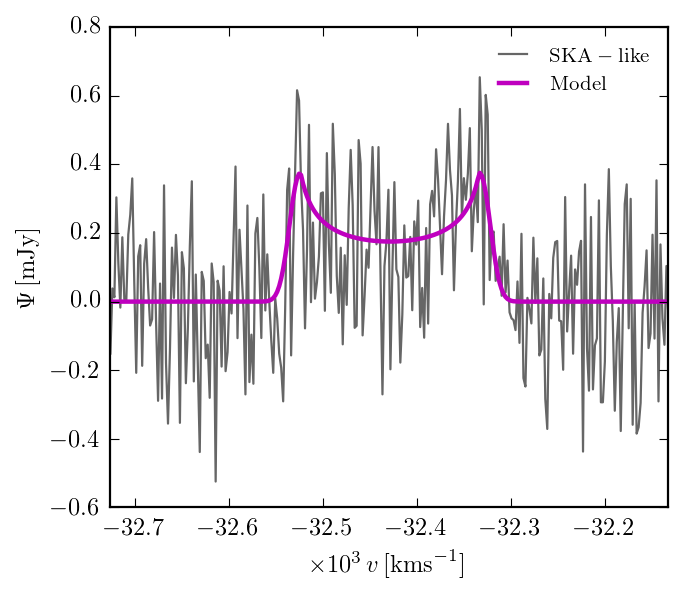}
\caption{Example \hi emission line data from our catalogue. Dark \ml{magenta} is the input double-horned profile and pale \ml{grey} is the simulated observation from an SKA1-MID experiment, where we have assumed the continuum has been subtracted and noise is Gaussian and uncorrelated between frequency channels. \emph{Left} shows the full simulated SKA Band 2 data and \emph{right} a zoomed in section around the line centre.}
\label{fig:example_line}
\end{figure*}
\subsection{Simulated catalogues}
In order to generate a realistic population of sources with \hi 21-cm line emission on which to try our technique, we make use of the set of \hi galaxy line profiles from the S3-SAX population as described in \cite{2009ApJ...703.1890O}, with intrinsic galaxy properties coming from a semi-analytic prescription for \hi emission painted on top of the Millennium N-body simulation \citep{2005Natur.435..629S}. In S3-SAX (which we will refer to as \skadssax for clarity), \hi 21-cm line emission is parametrised according to a simple six parameter, symmetric double-horn profile (e.g. the magenta line in \cref{fig:example_line}). The full \skadssax catalogue contains models for emission from some $3\E^{7}$ star-forming galaxy sources with \hi masses down to $1.85\E^{4} \, \mathrm{M}_{\odot}$, but here we are only interested in those sources which have been identified in a continuum observation and hence require redshift information. From a continuum galaxy sample we take the star-forming galaxies from the S3-SEX catalogue of \cite{wilman08} (which we will refer to as \skadssex) and rescale and cut them according to the requirements for weak lensing cosmology in SKA1-MID Band 2, as specified in \cite{2016arXiv160103948B}. This gives us a sample with a number density of resolved objects of $n_{\rm gal} = 2.7 \, \mathrm{arcmin}^{-2}$ and a median redshift $z_{\rm m} = 1.1$. We model this cut in the \hi catalogue from \skadssax as a redshift dependent cut on the \hi mass $M_{\rm HI}$:
\begin{equation}
\label{eqn:zcut}
M_{\rm HI} > z \times 10^{9.5}.
\end{equation}
This provides us with a sample with the same median redshift as the continuum $z_{\rm m} = 1.1$; the exact number density is not however replicated by this cut (\skadssax and \skadssex were constructed independently and do not contain matched objects). We have experimented with different versions of \cref{eqn:zcut} and found the most representative results (in terms of redshift and mass distributions) were obtained by matching the median redshifts. We therefore quote our results below as fractions of sources in \skadssax and normalise them relative to the numbers in \skadssex. For the sample selected, we expect the size (as given by \skadssax) to be $\sim 2 \pm 0.8 \, \mathrm{arcsec}$ with $\sim 0.03$ sources per beam at the lowest observing frequency (where the resolution will be $\sim 2 \, \mathrm{arcsec}$), meaning that some sources will be unresolved in the \hi observation but problems from confusion (i.e. multiple sources within the same beam) will be rare.

Previously, \cite{2015MNRAS.450.2251Y, 2015aska.confE..21S} define the signal-to-noise ratio (SNR) of each galaxy as:
\begin{equation}
\label{eqn:snr}
\mathrm{SNR}_{\rm vel} = \frac{v_{\mathrm{HI}}}{w_{\mathrm{peak}}^{\mathrm{obs}}}\frac{\sqrt{w_{\mathrm{peak}}^{\mathrm{obs}}/\delta V}}{S_{\mathrm{rms}}(\nu)},
\end{equation}
where $v_{\mathrm{HI}}$ is the velocity-integrated \hi line flux in Jy, $w_{\mathrm{peak}}^{\mathrm{obs}}$ is the width between the peaks of the double-horned profile in kms$^{-1}$ and $\delta V$ and $S_{\mathrm{rms}}$ (in Jy) are the frequency resolution and noise level of the experiment. They then cut the \skadssax sample accordingly and use these sources with $\snrvel$ and $\mathrm{SNR_{vel}} > 10$ as their detected samples of \hi galaxies to use in forecasting cosmological constraints. This is very much not a detection algorithm and hence does not model effects such as false detections or catastrophic failures, and it is interesting to see if the objects regarded as `detected' by this approach are replicated by our method.
\subsection{Experiments considered}
\label{sec:experiments}
We simulate the observation of the objects in the continuum detected population by a survey using the first phase of the SKA mid-frequency dish array (SKA1-MID). \cite{2015MNRAS.450.2251Y, 2015aska.confE..21S} specify \hi galaxy redshift surveys of $5,000 \degsq$ using $10,000$ hours of observing time in Band 1 and Band 2 of SKA1-MID. For each band we model the relative noise profile across the band from \cite{skabaseline} (see \cref{app:bands} for a full description), calculating $S^{\mathrm{ref}}_{\mathrm{rms}}$, the sensitivity at $\nu = 1 \, \mathrm{GHz}$ for comparison purposes (in the simulations we model the full frequency-dependent noise profile). The Band 2 survey is sensitive to sources with true redshift $0 < z < 0.58$ (corresponding to a frequency range $950 - 1420 \, $MHz) down to a reference noise level of $S^{\mathrm{ref}}_{\mathrm{rms}} = 187 \, \mu$Jy. The Band 1 survey has a frequency coverage of $350 - 1050 \, $MHz, corresponding to a redshift range of $0.35 < z < 3.06$ but with a higher noise level (driven mostly by the increase in sky temperature at lower frequencies) of $S^{\mathrm{ref}}_{\mathrm{rms}} = 315 \, \mu$Jy.  We assume $\delta \nu = 10 \,$kHz frequency channels covering this bandwidth, giving a total of 50,000 channels. As each data point corresponds to a correlation of different antenna pairs at different time points, the per-frequency channel noise can be modelled as uncorrelated and Gaussian with the relevant $S_{\mathrm{rms}}(\nu)$ \citep[see e.g.][]{1986isra.book.....T}. An example of the data sets considered can be seen in \cref{fig:example_line} which shows an SKA output both across the full Band 2 (with velocities given with respect to the rest frame velocity for the \hi 21-cm line) and zoomed in around the centre of the line profile.
\begin{table*}
\caption{Experiment parameters and summary of results for the simulated SKA observations. $S^{\mathrm{ref}}_{\mathrm{rms}}$ refers to the survey sensitivity at $\nu = 1 \, \mathrm{GHz}$ and is provided for heuristic comparison purposes only.}
\label{tab:results}
\begin{tabular}{l|c|c|c|c|c|c|c|c}
\hline
Experiment &  $z$ range & $S^{\mathrm{ref}}_{\mathrm{rms}}\,[\mu\mathrm{Jy}]$ & Selection & $f_{\mathrm{tot}}$ & $N_{\rm 5k}$ & $\eta_{\rm H}$ & $\eta_{\rm B}$ & $\eta_{3 \sigma}$ \\
\hline
\multirow{1}{*}{SKA1-MID Band 1} & \multirow{1}{*}{0.35 - 3.06} & \multirow{1}{*}{315} & 
$\ln(B) > 0$ & $0.18\%$ & $7.32 \times 10^{4}$ & 0.021 & 0.013 & 0.016 \\
\multirow{1}{*}{SKA1-MID Band 2} & \multirow{1}{*}{0.00 - 0.49} & \multirow{1}{*}{187} & 
$\ln(B) > 0$ & $10.14\%$ & $1.73 \times 10^{6}$ & 0.013 & 0.005 & 0.011 \\
\hline
\end{tabular}
\end{table*}
\section{Results and discussion}
\label{sec:results}
For each of the Band 1 and Band 2 surveys described in \cref{sec:experiments}, we find all of the lines in the input catalogue which have $\mathrm{SNR}_{\rm vel} > 1$ and estimate the posterior distributions for $\thetaline$ using \multinest. \ml{We assume that a line with $\mathrm{SNR}_{\rm vel} \leq 1$, is essentially undetectable and so do not fit these to save on computational resources.} \cref{fig:snr_b} strongly suggests that $B$ and $\mathrm{SNR}_{\rm vel}$ are correlated enough that we do not miss out on a large number of detectable lines. This fitting results in a full joint posterior distribution for all of the source parameters; here we present results for the redshift, derived from the estimate of the line centre velocity $v_0$ when marginalised over the other five parameters.

In summary, \cref{fig:band_meas} shows the numbers of detected sources (black line, right axes) and outliers (coloured lines, left axes) as a function of the sample cut on $B$, \cref{fig:band_1} shows two dimensional histograms of the recovered redshifts, which are shown in one dimension and compared to previous results in \cref{fig:band_zhist}. \cref{tab:results} gives numbers of redshifts available for the two experiments considered, along with outlier performance.
\subsection{Recovery rates}
Though we stress above the benefits of using the full posterior $P(z)$ and the Bayes Factor $B$ as inputs to downstream cosmological analyses as the best way to avoid biases, \ml{in order to evaluate our method}, here we present results with the estimated redshift $z_{\rm est}$ as the Maximum a Posteriori (MAP) redshift from the $P(z)$ and applying a cut to our sample based on the $B$, in order to remove spurious fits to noise features.

We define $N_{\rm band}$ as the number of input sources from the \skadssax catalogue which have their \hi 21-cm emission line redshifted into the relevant observing band. Using this cut and $z_{\rm est}$ we calculate the $f_{\rm tot}$ fraction of these sources which have their redshift recovered by each method. We also calculate the number of sources these fractions correspond to when applied to the fiducial $5000 \, \deg^{2}$ continuum cosmology survey:
\begin{equation}
N_{\rm 5k} = f_{\rm tot}N^{\rm cont}_{\rm band},
\end{equation}
where $N^{\rm cont}_{\rm band}$ is the number of sources in the continuum survey which have \hi 21-cm lines redshifted into the relevant observing band. This $N_{\rm 5k}$ is then the number of redshifts a given method may be expected to provide.
\subsection{Catastrophic outliers}
\label{sec:zest}
From the full posterior $P(z)$ distribution for each source, we calculate a number of metrics for redshift quality, quantifying the distance between $z_{\rm est}$ and the true redshift $z_{\rm true}$. Two are `catastrophic outlier fractions':
\begin{equation}
\eta = N_{\rm out} / N_{\rm band},
\end{equation}
where $N_{\rm band}$ is the total number of sources which have lines redshifted into the relevant observing band, and $N_{\rm out}$ is the number of these sources with redshift estimates far enough from the true redshift to be classified as outliers.

\begin{figure*}
\includegraphics[width=0.475\textwidth]{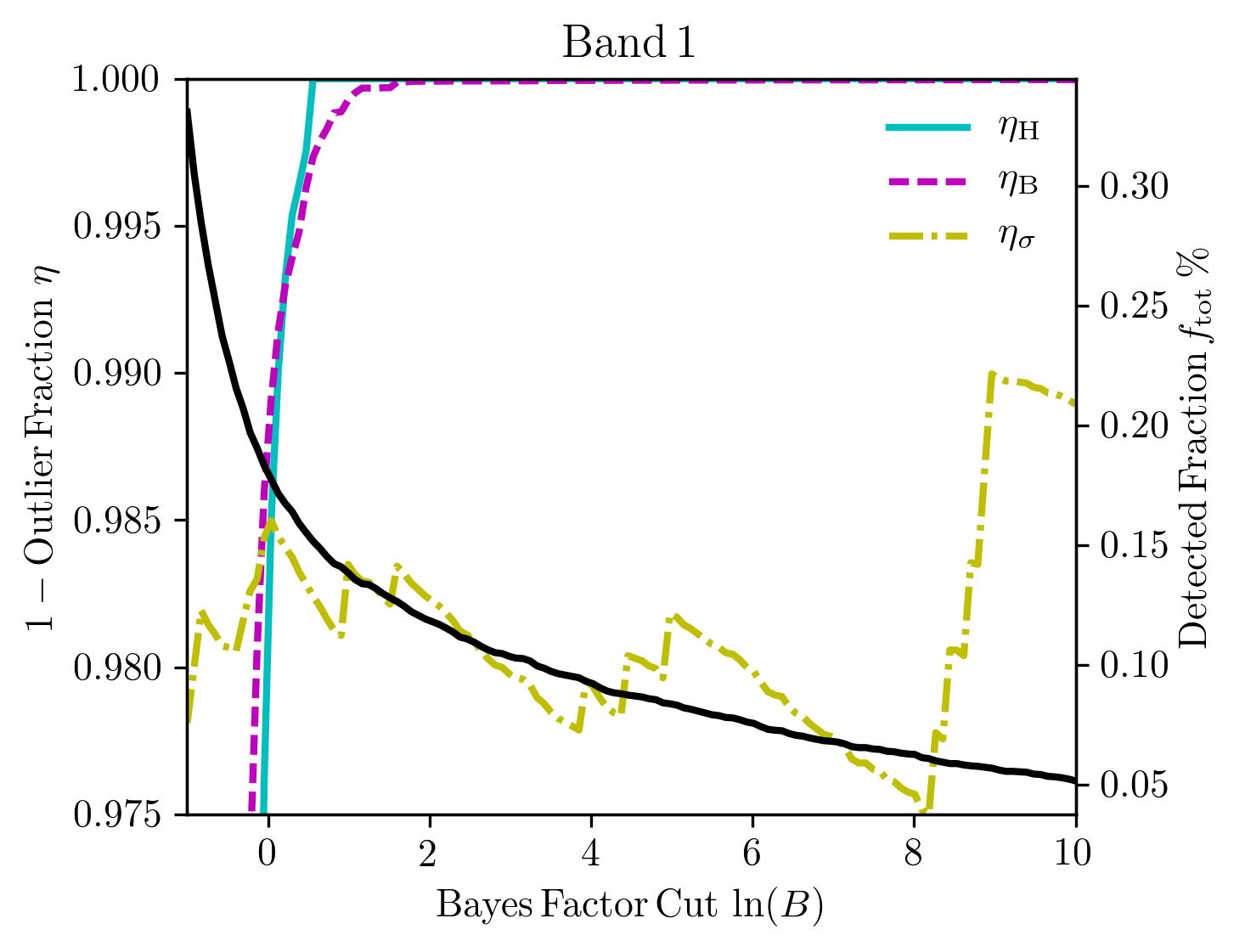}
\includegraphics[width=0.475\textwidth]{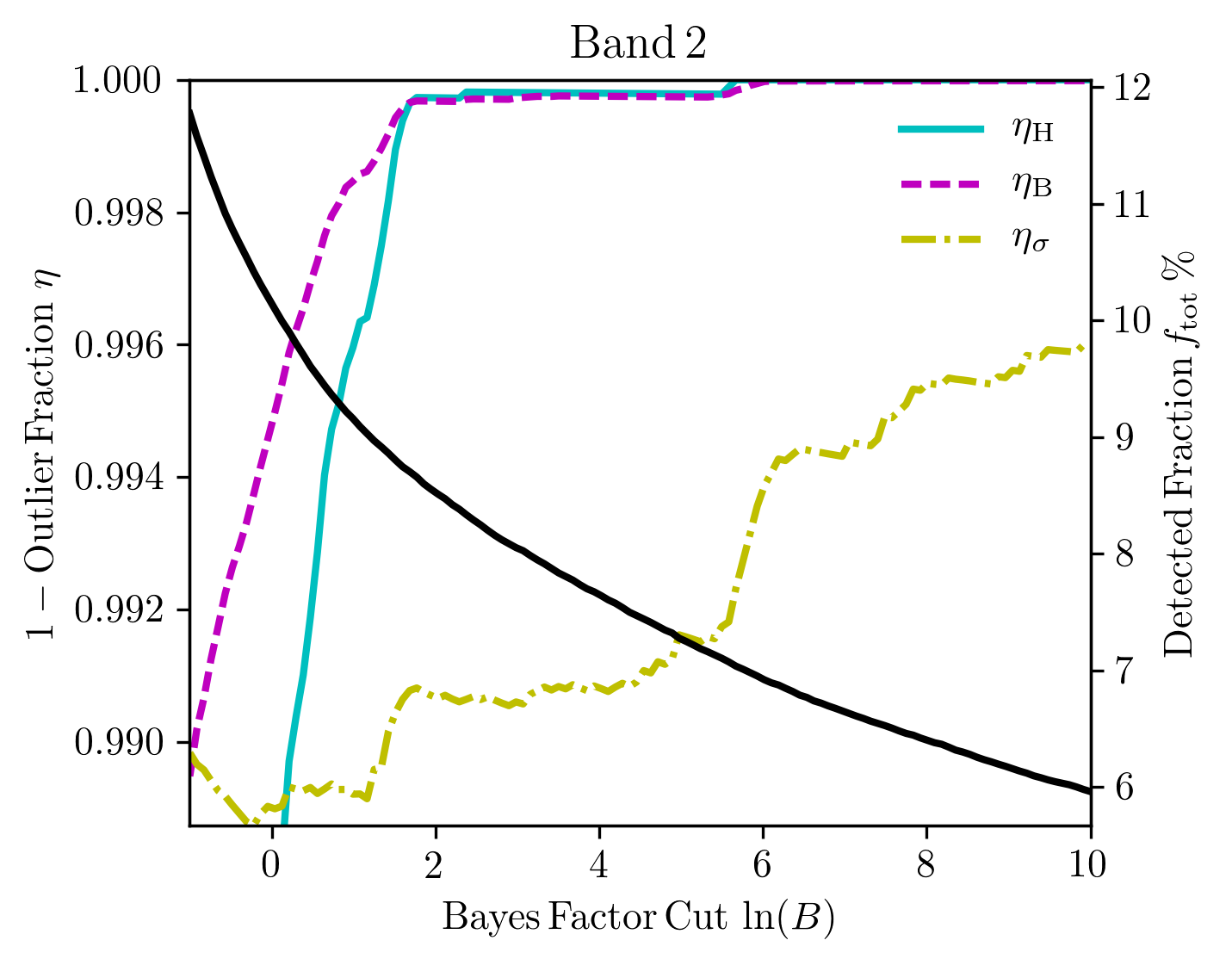}
\caption{Fractions of line estimates which are far enough from the true redshift to be classified as outliers (coloured lines, left axes, with the three different outlier definitions from \cref{sec:zest}) and detected fractions (bold black line, right axes) of objects with a line redshifted into the band for the Band 1 and Band 2 surveys, as a function of the Bayes factor cut applied to the fitted sample. We adopt a fiducial cut of $\lnb$ in the results presented below.}
\label{fig:band_meas}
\end{figure*}

We use two classifications of outliers, with the first given by Table 2 of \cite{2015MNRAS.449.1275H} as the number of redshift estimates with:
\begin{equation}
\left| \frac{z_{\rm est} - z_{\rm true}}{1 + z_{\rm true}} \right| > 0.15,
\end{equation}
the outlier fraction for which we refer to as $\eta_{\rm H}$. The second is given by \cite{2010MNRAS.401.1399B} as redshift estimates with:
\begin{equation}
\left| \ln \frac{1 + z_{\rm est}}{1 + z_{\rm true}} \right| > 0.2,
\end{equation}
the outlier fraction for which we refer to as $\eta_{\rm B}$. Typical catastrophic outlier fractions defined in this way for photometric redshift estimators using optical and near-IR data are $\sim 2$ per cent and outlier fractions for our method are shown as a function of the $B$ cut in \cref{fig:band_meas}. We also show in \cref{tab:results} the fraction $\eta_{3\sigma}$, for which $N_{\rm out}$ is given by the number of lines for which $z_{\rm true}$ is outside of the Credible Interval containing $99$ per cent of the posterior probability mass, equivalent to being outside of the $3 \sigma$ region of a Gaussian likelihood (the appearance of this line in \cref{fig:band_meas} is due to low absolute numbers of sources making up the curves). \cref{fig:band_meas} shows the completeness (detected fraction, right axes) and purity ($1 - \eta$, left axes) for the Band 1 and 2 surveys described above as a function of the Bayes factor $B$ used to cut the sample. As can be seen, detection fractions $f_{\rm tot}$, are relatively low at $\sim0.25$ per cent for Band 1 and $\sim10$ per cent for Band 2, reflecting the low sensitivity of SKA1-MID to the relatively faint \hi signal in these redshift ranges. We chose $\ln(B) > 0$ as a fiducial cut, and present results below for this sample.

We also perform 1000 realisations of Band 1 and 1000 realisations of Band 2 data containing no signal and only noise, and again attempt to fit a six parameter double horn profile to the data. This allows us to quantify the effect of spurious detections when continuum detected galaxies have their 21-cm line redshifted outside of the relevant observing band. We find that, out of the 2000 noise-only realisations simulated across both bands, only 22 (all in Band 2) have a value for the Bayesian evidence favouring the signal model over the noise model, with our fiducial cut of $\lnb$.

\subsection{Redshift estimates}
\begin{figure}
\includegraphics[width=0.475\textwidth]{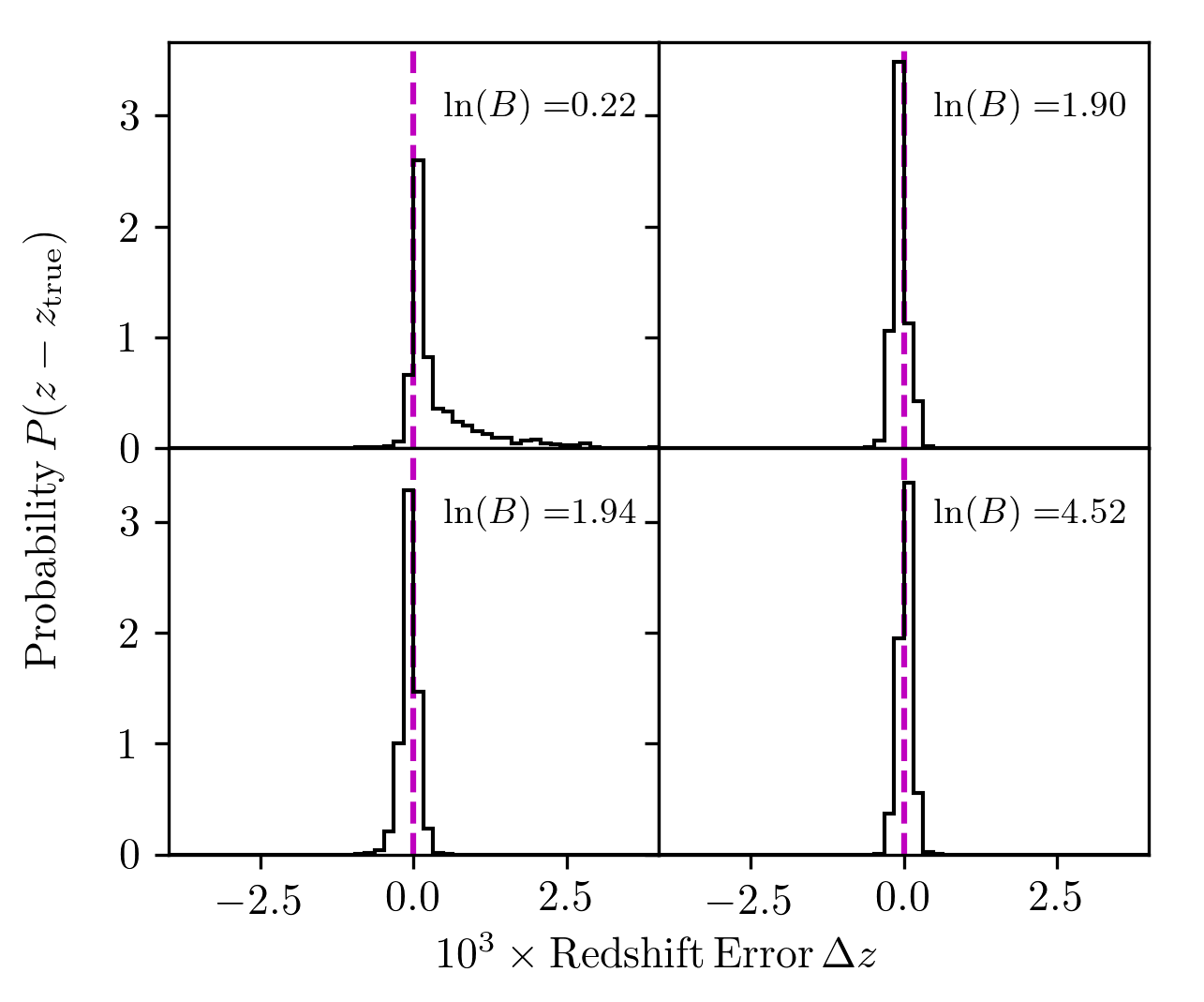}
\caption{Example $P(z)$ for four lines within our catalogue, showing the good performance with small $\mathcal{O}(10^{-3})$ redshift uncertainties even for low Bayes factors.}
\label{fig:pz_examples}
\end{figure}

\begin{figure*}
\includegraphics[width=0.475\textwidth]{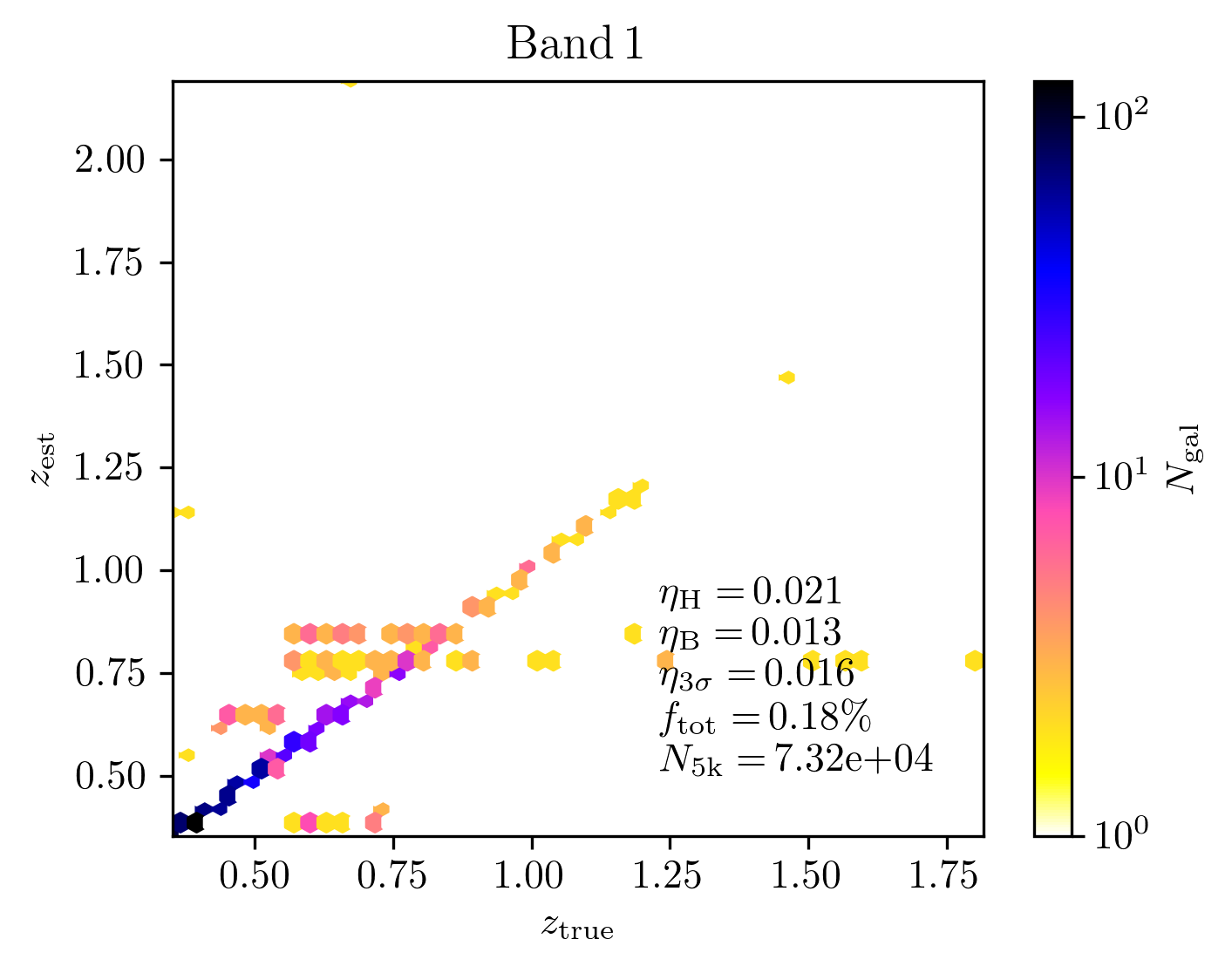}
\includegraphics[width=0.475\textwidth]{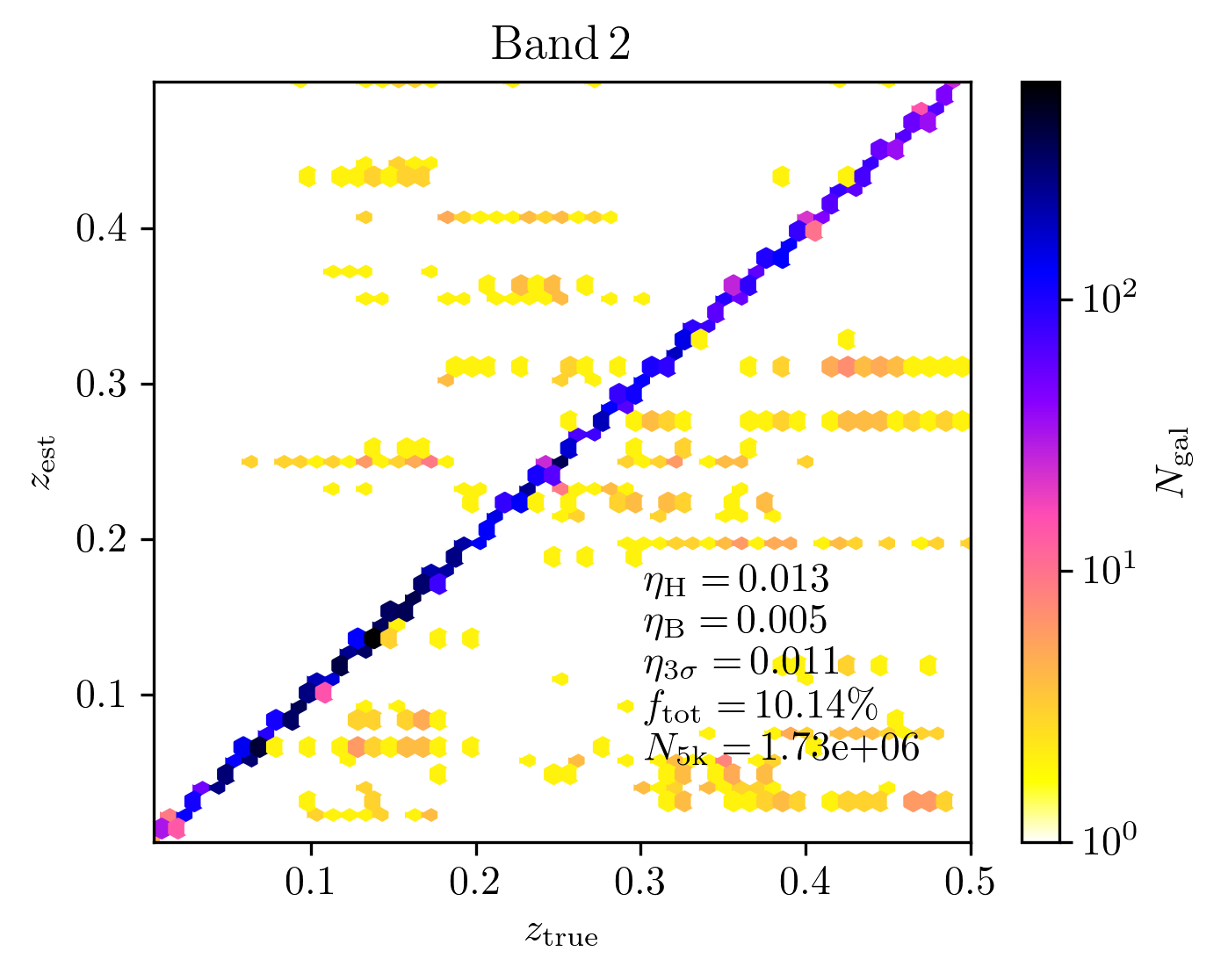}
\caption{Two dimensional histogram of estimated vs true redshift for Band 1 (\emph{left}) and Band 2 (\emph{right}) of SKA1-MID, showing the differing recovered redshift distributions and presence of catastrophic outliers. $\eta$ are outlier fractions as defined in the text, $f_{\rm tot}$ is the fraction of galaxies with lines redshifted into the band which have a redshift recovered and $N_{\rm 5k}$ is the number of continuum galaxies in a $5000 \, \deg^2$ survey this would represent.}
\label{fig:band_1}
\end{figure*}

\begin{figure*}
\includegraphics[width=0.475\textwidth]{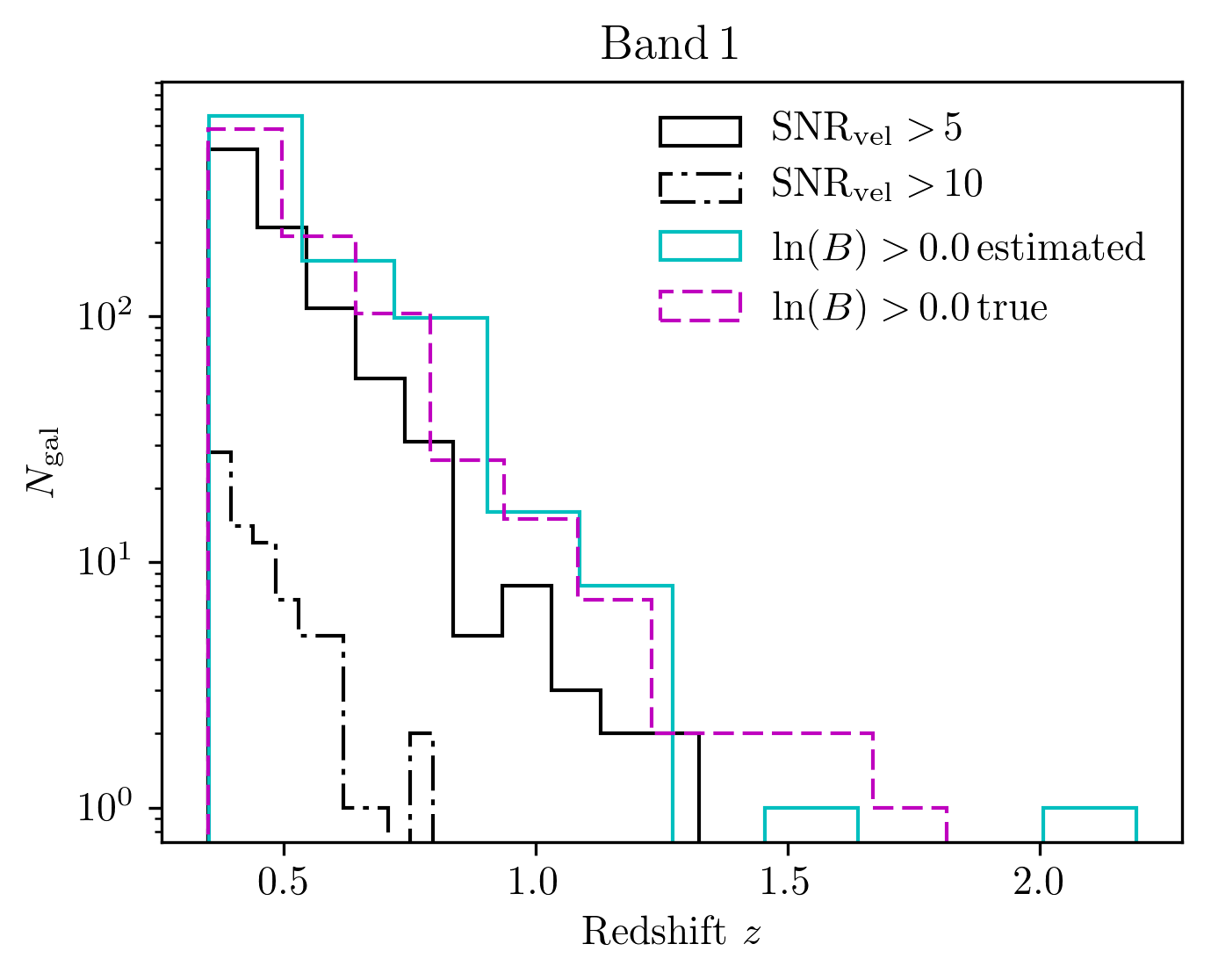}
\includegraphics[width=0.475\textwidth]{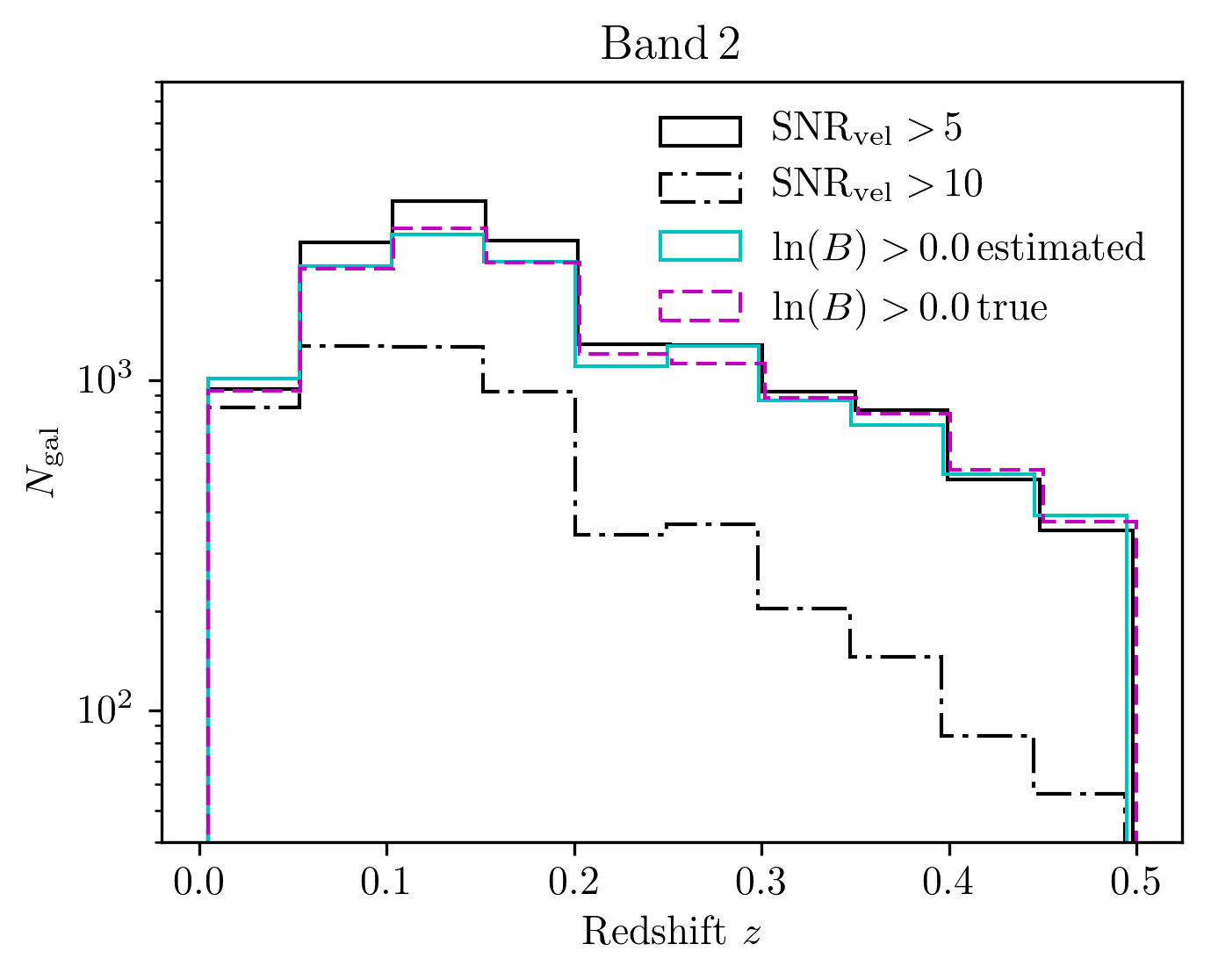}
\caption{Redshift distributions of the $\lnb$ and $\snrvel$ samples, along with the true redshift distribution for the $\lnb$ sample, showing the good recovery of the input distributions.}
\label{fig:band_zhist}
\end{figure*}

The left and right panels of \cref{fig:band_1} shows two dimensional histograms of $z_{\rm est}$ against $z_{\rm true}$ for the $\ln(B) > 0$ sample in SKA1-MID Bands 1 and 2 respectively, with the summary statistics also presented in \cref{tab:results} and four examples of marginalised $P(z)$ distributions shown in \cref{fig:pz_examples}. Good redshift recovery can be seen for the majority of sources included after the $\lnb$ cut, with relatively few outliers. Redshifts are well recovered across Band 2, but are extremely few above $z\sim1$ in Band 1. In \cref{fig:band_zhist} we show one dimensional histograms of the estimated redshifts of the sources recovered by our method with the $\lnb$ cut, along with the true redshift distribution of these sources, showing excellent agreement. For reference, we also cut our simulated sample with $\snrvel$ and $\mathrm{SNR}_{\rm vel} > 10$ as performed in \cite{2015MNRAS.450.2251Y} and \cite{2015aska.confE..21S} respectively for their cosmological forecasts and display the redshift range of the resulting sample. The $\lnb$ and $\snrvel$ samples share highly similar redshift distributions, even if the exact set of objects selected by the two cuts are not the same, as discussed in \cref{sec:snr_comp}.
It should be noted that the \cite{2015MNRAS.450.2251Y} and \cite{2015aska.confE..21S} studies were performed before the re-baselining of SKA1-MID, which resulted in significant changes to the sensitivity curves of both Band 1 and Band 2, meaning they are not directly comparable to the samples presented there \citep[though see][for forecasts using the $\mathrm{SNR_{vel}}$ detection method with the re-baselined telescope]{2016ApJ...817...26B}. These previous studies also allowed \hi line emission to be found for all sources in the \skadssax sample, without requiring the continuum detection we implement via \cref{eqn:zcut}, which would require blind line finding in the full spatial and spectral resolution data cubes emanating from the telescope. If continuum selection is used, we estimate data volumes may be decreased from $100 \, \mathrm{TB}$ for a single pointing ($3600\times3600$ spatial pixels covering $1 \deg^2$ over $10000$ frequency channels) for the full cube to $0.075 \, \mathrm{TB}$ per pointing for one dimensional, $10000$ frequency channel data vectors for all of the $2.7$ galaxies per square arcminute.

\subsection{Relation to SNR definitions}
\label{sec:snr_comp}
\begin{figure*}
\includegraphics[width=0.475\textwidth]{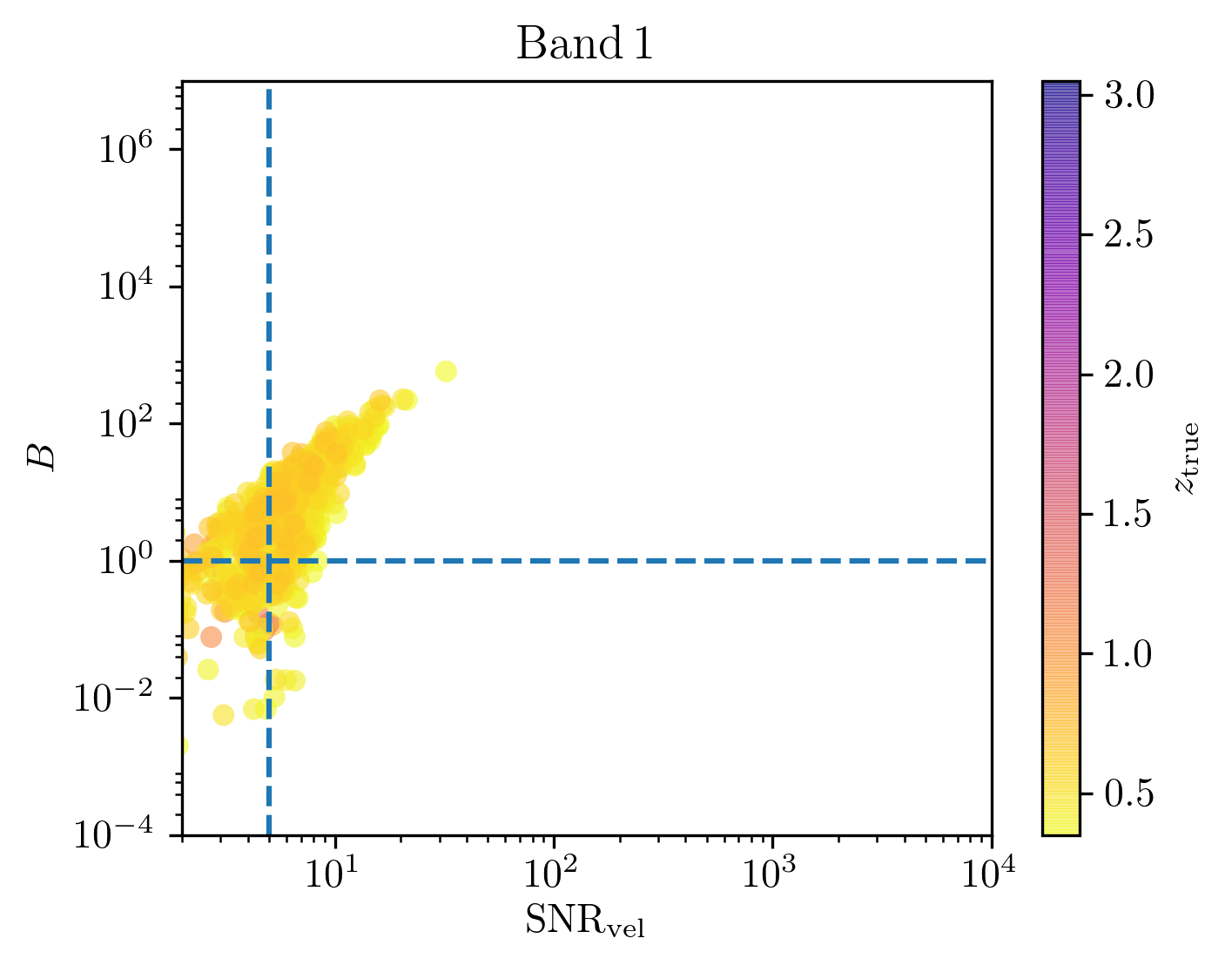}
\includegraphics[width=0.475\textwidth]{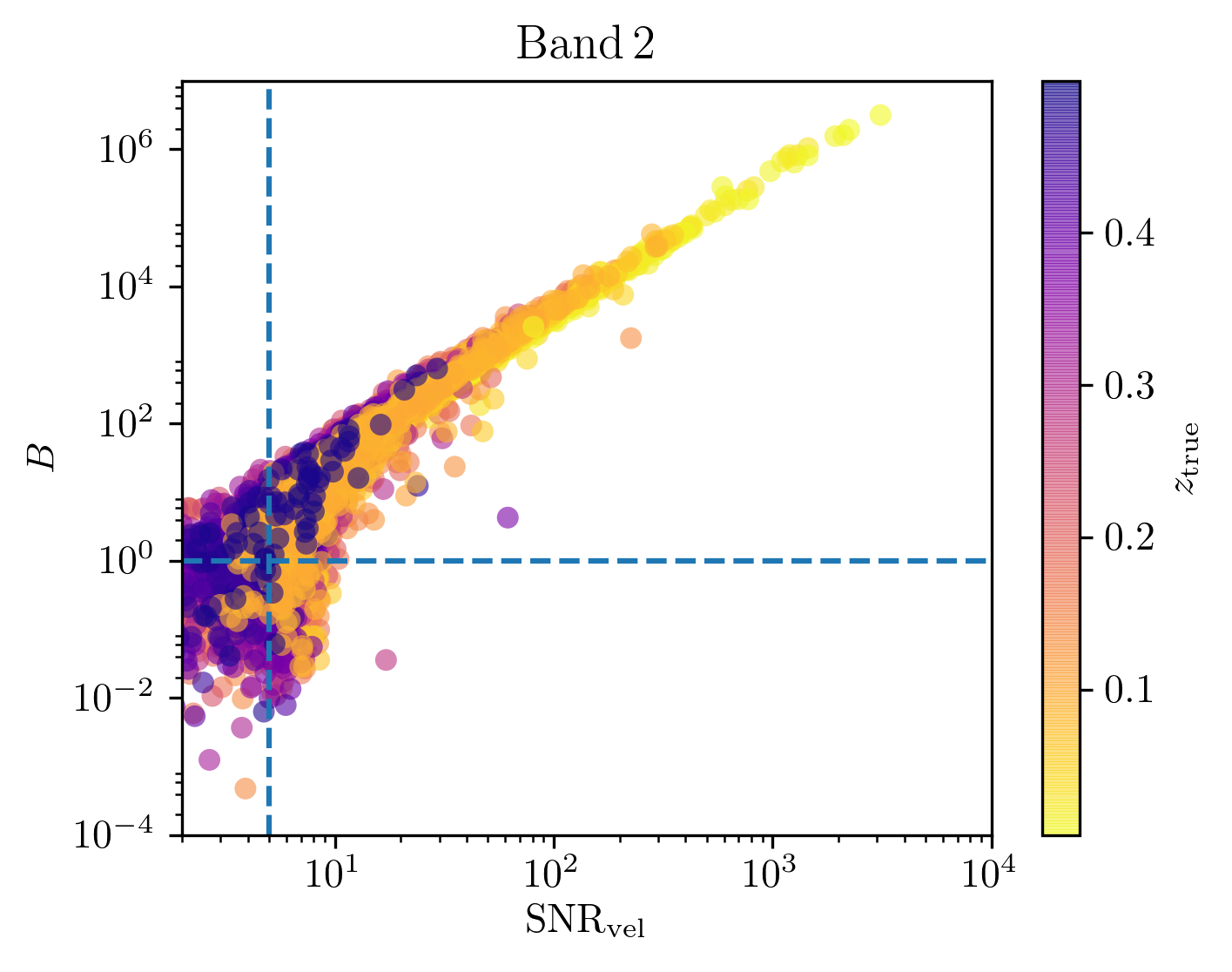}
\caption{The Bayesian evidence ratio for true detection of the presence of a line $B$ and the $\mathrm{SNR_{vel}}$ (Equation \ref{eqn:snr}) for each source in Band 1 (\emph{left}) and Band 2 (\emph{right}). The colour scale shows the true redshift of the sources, and the horizontal and vertical lines represent the fiducial $\lnb$ and $\snrvel$ cuts respectively. Note that there are populations of sources included that pass the SNR cut, but are not detected by our method, even if the redshift distributions in \cref{fig:band_zhist} appear the same, and that only sources passing the continuum detection threshold \cref{eqn:zcut} and with $\mathrm{SNR_{vel}} > 1$ are shown. The Bayesian method presented in this paper is a true  detection method and so should more closely reflect the population available in SKA1-MID surveys. See \cref{fig:snr5_lines} for an illustration of how lines with the same SNR may differ in whether or not they are detected.}
\label{fig:snr_b}
\end{figure*}

\begin{figure}
\includegraphics[width=0.475\textwidth]{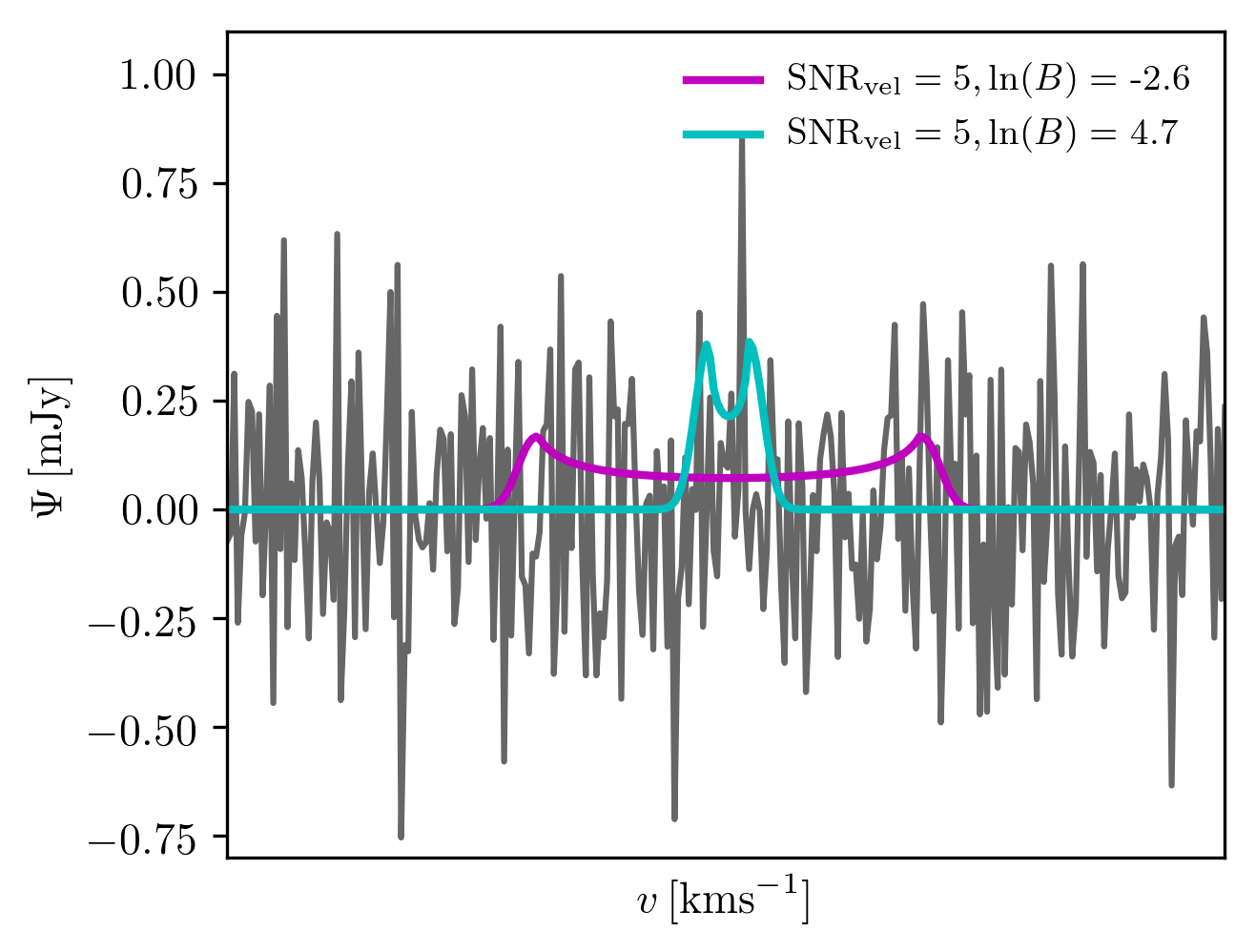}
\caption{Example \hi emission line data from our catalogue, with input line shapes plotted with representative SKA1-MID noise. Both lines in this figure have $\mathrm{SNR_{vel}}=5$ but the narrower, taller line is classed as detectable by our method but the shorter, wider line is not.}
\label{fig:snr5_lines}
\end{figure}

\cref{fig:snr_b} shows the relation between two measures of significance of line redshift detection: both the recovered Bayes factor $B$ and the $\mathrm{SNR}_{\rm vel}$ definition. Whilst a correlation can be seen between the two measures of detection significance, there are important differences. The two dashed lines mark the fiducial cuts for the two approaches: the vertical line for $SNR_{\rm vel} > 5$ and the horizontal line for $\ln(B) > 0$. These two cuts select two different populations in detail; the upper right quandrant is selected by both methods, the upper left by only the $\ln(B) > 0$ cut, the lower right by only the $\mathrm{SNR_{vel}}>5$ and the lower left quadrant by neither. \ml{We note that with a real \hi line detection technique, an SNR cut even of 10 does not guarantee detection nor preclude detection of galaxies well below the threshold.} In \cref{fig:snr5_lines} we show two examples of $\mathrm{SNR_{vel}}=5$ lines with representative noise for the Band 2 survey. Whilst both of these lines have the same $\mathrm{SNR_{vel}}$, the narrower, taller line is significantly detected, with $\ln(B) = 4.7$ whilst the shorter, broader line is not, having only $\ln(B) = -2.6$. 

We stress that the results presented here are highly robust, coming from a full simulation of the relevant data and actual application of the detection methods, rather than simply calculating \cref{eqn:snr} for the input catalogue and applying a cut, a procedure which will not model the real recovery rate and issues such as outliers and false detections. We therefore believe that the sample selected by $\ln(B) > 0$ is much closer to the set of galaxies with \hi redshifts which will be truly detectable with SKA1-MID.
\section{Extensions and improvements}
\label{sec:extensions}
The method we have presented here fits a six parameter model with broad, uninformative priors to the one dimensional data sets considered. However, a less conservative and more constraining approach changing the priors in \cref{tab:priors} to be more informative could be well-motivated. On a catalogue level, the results of previous surveys could be used as priors on e.g. the redshift distribution and luminosity functions of the sources, taking the form of non-uniform priors on $v_0$ and $\Psi^{\rm obs}$ parameters.

More informative priors could also be used individually for each source from auxiliary data. In particular, the continuum size, shape, flux, orientation and velocity dispersion are all correlated with, and can be used to form useful priors on, the parameters describing the \hi line profile for the source. For resolved sources, morphology information could also be used; for example in constraining more parameters in the line profiles such as asymmetric heights between the two horn peaks (as may be expected for sources with irregular shapes). Model extensions could also allow for mitigation of Radio Frequency Interference (RFI) contamination of the spectra, for example by allowing Bayesian model comparison between RFI feature models and the emission line profile.

Further extensions could also be made to the six parameter model, allowing for inclusion of an un-subtracted continuum flux or more complicated emission line profiles \citep[e.g. the Busy function of][]{2014MNRAS.438.1176W} using only a small number of extra parameters in the fit.

Additionally, in the event of confused sources, Bayesian model selection could be used to determine the optimal number of sources and disentangle the confused signals.

The choice of radio \ml{continuum} sources as \ml{the source of position information} is not unique. For instance, one could instead choose position priors from surveys at optical wavelengths, such as \emph{Euclid} or LSST, and perform emission line fits to the data at these locations, potentially providing spectroscopic-like redshifts from the radio data. This has the potential to be extremely effective with the advent of the full SKA (SKA2), which is expected to have up to $10$ times the sensitivity of SKA1 -- the optimistic scenario for SKA2 in \cite{2015MNRAS.450.2251Y} containing the same numbers of spectroscopic sources as a \emph{Euclid} imaging survey up to $z\sim2$.

As well as source redshifts, our results can also provide full posterior probability distributions for other parameters which are derived from the shape of the emission line, such as the \hi mass of galaxies -- a highly important quantity in studies of the star formation and galaxy evolution across cosmic time.
\section{Conclusions}
\label{sec:conclusions}
We have investigated the ability of Bayesian model fitting of \hi 21-cm line emission to provide redshifts for galaxies in radio continuum surveys. We use the continuum detection of a galaxy \ml{to provide} its sky location and fit a six parameter model to the resulting one dimensional data set.

When this redshift estimation method is applied to simulations of SKA1-MID observations and a cut, on the Bayesian evidence ratio between signal and noise models, of $\lnb$ applied, we find that we may recover redshifts with good confidence $\lnb$ for up to $10.14$ per cent of low-redshift objects using observing Band 2, and up to $0.18$ per cent of high-redshift objects using observing Band 1. When compared to the previous $\snrvel$ selection of \cite{2015MNRAS.450.2251Y} we recover similar objects in terms of their number and redshift distribution, but our selection represents a significant improvement in sophistication: performing full data level simulations and attempting to recover the line profiles, rather than simply calculating an estimated SNR from simulated intrinsic source properties. Our method also has the significant advantage of providing a full $P(z)$ and detection significance $B$ for each source, which may be coherently folded in to cosmological parameter estimation analyses and other analyses such as estimation of galaxy formation histories via the \hi mass function.

This allows a firm quantification to be made on the numbers of redshifts which will be available for continuum selected sources from SKA1-MID using only the radio data, giving insight into the extent of the reliance on cross-matched catalogues in other wavebands to obtain source redshifts. This should place current SKA cosmology forecasts on firmer footing and potentially allow for improvement. The performance of our redshift estimator could also inform the design of SKA data processing strategies, potentially allowing larger bandwidths and higher frequency resolutions, since the use of a continuum prior on source location obviates the need for blind finding of sources in extremely large three dimensional (spatial and spectral) data cubes.

Here we have only considered the improvement for the first phase of the SKA (SKA1) but the full SKA (SKA2) should present even more opportunities. With the correct bandwidth, large and fast surveys with SKA2 could potentially become a redshift machine providing posterior $P(z)$ for \emph{Euclid} and LSST sources (the continuum prior information need not come from a radio survey), circumventing many of the systematics of photometric redshifts which may otherwise limit their cosmological constraints.

\section*{Author Contributions}
IH was responsible for the initial conception of the project. IH and ML designed the study, developed the methodology, performed the analysis, and wrote the manuscript. ML wrote the analysis code and ran the simulations. MLB contributed to the definition of the galaxy samples and to the initial development of the simulations.

\section*{Acknowledgments}
We would like to thank Adam Avison, Bruce Bassett, Anna Bonaldi, Phil Bull, Stefano Camera, Keith Grainge, Mario Santos and Joe Zuntz for helpful discussions and comments on the draft. We also thank Robert Braun and Phil Bull for providing us with the SKA sensitivity curves and Hiranya Peiris for allowing access to the Hypatia cluster at UCL. IH is supported by an ERC Starting Grant (grant no. 280127). ML acknowledges support from the SKA, NRF and AIMS. This work is partially supported by the European Research Council under the European Community's Seventh Framework Programme (FP7/2007-2013)/ERC grant agreement no 306478-CosmicDawn.


\appendix
\section{SKA1-MID Noise Profiles}
\label{app:bands}
\begin{figure}
\includegraphics[width=0.475\textwidth]{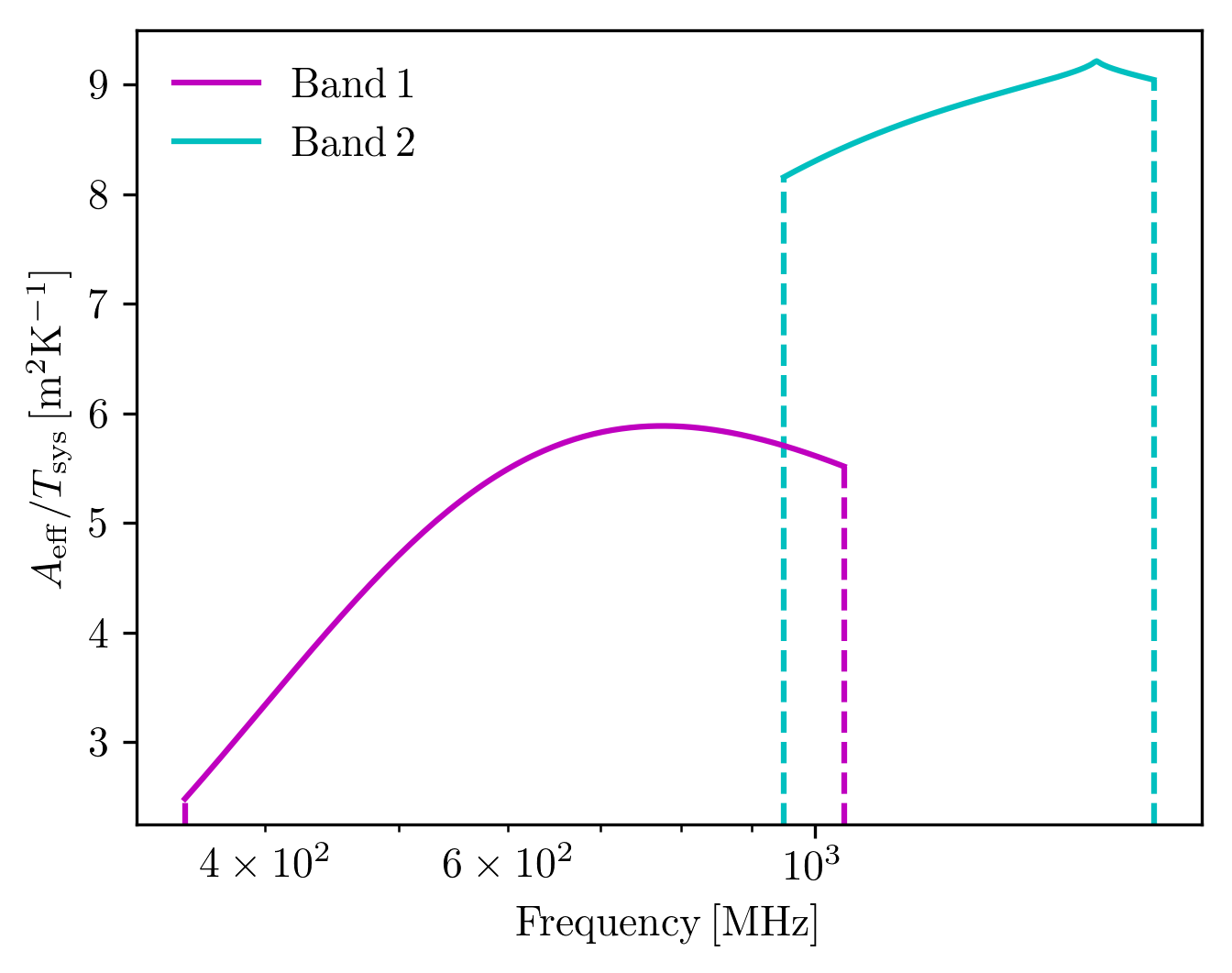}
\caption{The $A_{\rm eff}/T_{\rm sys}$ for the two SKA1-MID observing bands used in this paper.}
\label{fig:band_noise}
\end{figure}

The mid-frequency dish array of phase 1 of the Square Kilometre Array (SKA1-MID) is currently still under design, meaning the exact noise properties of the telescope are still somewhat uncertain. For this work we make use of the most recent noise curves publicly available \citep[Figure 8 of][with equations provided by Robert Braun in private communication]{skabaseline}, on the understanding they are likely to be representative of the true performance. We model the receiver temperatures (in Kelvin) for SKA1-MID Band 1 and Band 2 with frequency ($\nu$ in GHz) dependencies:
\begin{equation}
T^{\rm B1}_{\rm rcv} = 11 + 3\left(\frac{\nu - 0.35}{1.05 - 0.35}\right)
\end{equation}
and:
\begin{equation}
T^{\rm B2}_{\rm rcv} = 8.2 + 0.7\left(\frac{\nu - 0.95}{1.75 - 0.95}\right).
\end{equation}
To calculate the system temperature we also use the sky temperature:
\begin{align}
T_{\rm sky} =& 20 \left( \frac{0.408}{\nu}\right)^{2.75} + 2.73 \nonumber \\ &+ 288\left[ 0.005 + 0.1314 \exp(8 \times 10^{\nu - 22.23}) \right],
\end{align}
spillover temperature:
\begin{equation}
T_{\rm spl} = 4
\end{equation}
and the ground temperature:
\begin{equation}
T_{\rm gnd} = 300,
\end{equation}
with $T_{\rm sys} = T_{\rm rcv} + T_{\rm sky} + T_{\rm spl}$. We also calculate the antenna efficiency as a function of frequency, from the dish diameter $D_{\rm dsh}$:
\begin{align}
\eta = \eta_{0} - 70\left( \frac{c}{D_{\rm dsh} \nu \times 10^{9}} \right)^{2} - 0.36\left( \frac{\left| \nu - 1.6 \right|}{24 - 1.6} \right)^{0.6}
\end{align}
giving the effective area for the combined number of $N_{\rm ant}$ dishes:
\begin{equation}
A_{\rm eff} = N_{\rm ant} \eta \pi 0.25 D_{\rm dsh}^{2}
\end{equation}
and the resultant System Equivalent Flux Density:
\begin{equation}
SEFD = 2k_{\rm B} \frac{T_{\rm sys}}{A_{\rm eff}}.
\end{equation}
\cref{fig:band_noise} shows the resultant $A_{\rm eff} / T_{\rm sys}$ from this recipe, with $N_{\rm ant} = 190$ and $D_{\rm dsh} = 15 \, \mathrm{m}$ for the two SKA1-MID observing bands considered in the paper. These may then be used to calculate a noise rms:
\begin{equation}
S_{\rm rms} = 260 \, \mu\mathrm{Jy} \left(\frac{T_{\rm sys}}{20 \, \mathrm{K}}\right) \left(\frac{25,000 \, \mathrm{m}^2}{A_{\rm eff}}\right) \left(\frac{0.01 \, \mathrm{MHz}}{\delta\nu}\right)^{1/2} \left(\frac{1 \, \mathrm{h}}{t_{\rm p}}\right)^{1/2}
\end{equation}
where $\delta\nu$ is the frequency channel width and $t_{\rm p}$ the pointing time \ml{(which we assume to be 1.76 hours as in \citealt{2015MNRAS.450.2251Y}}).

\bsp	
\label{lastpage}
\end{document}